\documentclass[12pt,preprint]{aastex}

\newcommand \etal{{\em et al.}}

\shorttitle{Radio Continuum and Star Formation in CO-rich Early Type Galaxies}
\shortauthors{Lucero \etal.}
\begin{document}
\title{Radio Continuum and Star Formation in CO-rich Early Type Galaxies}
\author{Danielle M. Lucero and Lisa M. Young}
\affil{Physics Department, New Mexico Institute of mining and Technology,
Socorro, NM 87801} 
\email{drundle@nmt.edu, lyoung@aoc.nrao.edu}
\nocite{*}
\begin{abstract}

In this paper we present new high resolution VLA 1.4 GHz radio continuum observations of five FIR bright CO-rich early-type galaxies and two dwarf early-type galaxies.  The position on the radio-FIR correlation combined with striking agreements in morphology between high resolution CO and radio maps show that the radio continuum is associated with star formation in at least four of the eight galaxies. The average star formation rate for the sample galaxies detected in radio is $\sim$2 M$_\odot$ year$^{-1}$.  There is no evidence of a luminous AGN in any of our sample galaxies.  We estimate Toomre $Q$ values and find that the gas disks may well be gravitationally unstable, consistent with the above evidence for star formation activity.  The radio continuum emission thus corroborates other recent suggestions that star formation in early type galaxies may not be uncommon.

\end{abstract} 
\keywords{galaxies: elliptical and lenticular, cD --- galaxies: ISM --- radio continuum: ISM --- infrared: ISM --- stars: formation}
\section{Introduction}
It is well established that a number of early-type galaxies contain small amounts of cold interstellar gas and dust (eg. Knapp \etal~1989; Colbert, Mulchaey \& Zabludoff 2001; Sadler \etal~2000) presumably obtained through mergers/accretion or stellar mass loss.  What is not well known is the ultimate fate of this cold gas and dust.  van Gorkom et al. (1989) estimate that several hundredths of a solar mass per year may be falling onto and feeding the nuclear black hole in some early-type galaxies.  It is also possible that the cold gas may be destroyed due to the presence of a hostile environment.  On the other hand, if the neutral gas does not get stripped, destroyed, or fall into a central AGN, it may settle into a disk and form stars.  It has been argued by Okuda \etal~(2005) that gas disks in CO-rich early-types are stabilized by their large dynamical masses, and so star formation is much less likely to occur in these galaxies than in late-type galaxies.  However, many authors (e.g. Kauffman 1996; Scorza \& Bender 1995, 1998) agree that if a significant episode of star formation does occur in an early-type galaxy, then it may be transformed into something resembling a late-type galaxy.  Indeed, many early-types are known to contain small stellar disks at their centers (Scorza \& Bender 1998; Emsellem \etal~2004).  The possibility of star formation in early-type galaxies is important since it challenges the long held belief that early-type galaxies are passive evolvers. 

Recently, Kaviraj \etal (2006) used the Galaxy Evolution explorer (GALEX) to observe a large sample of low redshift early-type galaxies in the near ultraviolet (NUV).  They combine these observations with optical data from the Sloan Digital Sky Survey (SDSS) and find that at least $\sim$30 percent of the sample galaxies have UV to optical colors consistent with some recent ($<$Gyr) star formation.  Their findings are in agreement with semi-analytical $\Lambda$CDM hierarchical merger models which predict that low redshift early-types will contain $\sim$1 to 3 percent of their total stellar mass in stars less than 1 Gyr old.  They also determine that the observed blue colors in the star forming fraction of their sample cannot be produced by assuming a ``monolithic'' model of evolution.  Furthermore, their results show that star formation is equally likely to occur in either high density or field environments.  Although these results strongly suggest that a large fraction of early-type galaxies are currently forming stars, no conclusions can be drawn as to the actual star formation rates (which could show some environmental dependence).    

Whether a galaxy is forming stars is sometimes traced by its placement
 on the FIR-radio correlation known to exist in normal spiral galaxies
 (Young \& Scoville 1991; Condon 1992).  The radio continuum emission
 in these galaxies arises in HII regions, type II supernovae and the
 synchrotron emission from the electrons accelerated by those
 supernovae.  Dust heated by these same massive stars is thought to
 produce the FIR emission from spiral galaxies. Studies of
 interstellar medium of early-type galaxies have shown that their cold
 gas is physically similar to that of spirals (Rupen 1997; Young \& Lo
 1996; Young 2002).  Based on this, it is easy to imagine that star
 formation might be traced in early-types in the same manner as normal
 spiral galaxies.  However,  many early-types contain AGN, and so they may
 have multiple sources of radio continuum.  Clearly, the total radio
 fluxes combined with a knowledge of the radio morphology are needed
 in order to disentangle these sources of radio emission. 

 Wrobel \& Heeschen (1988) were the first to suggest that several early-type galaxies might be undergoing significant star formation, based on both placement on the FIR-radio correlation and the fact that their 6 cm continuum detections are extended on scales of 10 kpc or more.  Cotton and Condon (1998) find support for this idea in the radio and FIR flux ratios of E/S0s detected in the NRAO VLA Sky Survey (NVSS).  About a quarter of them have flux ratios consistent with star formation activity.  The other three quarters are presumably powered by AGN.  

While studies of this kind can rule out the presence of a luminous AGN, they cannot definitively prove that the radio continuum is associated with recent star formation for several reasons.  Firstly, early-type galaxies are fundamentally different from spirals, with different stellar populations and different stellar dynamics, hence different types of energy input into the ISM.  Secondly, there is evidence to suggest that post Asymptotic Giant Branch (AGB) stars can provide enough UV radiation to heat the dust and ionize the gas and account for the observed IR luminosities (Ferrari \etal~1999).  In other words, perhaps both the IR and the radio continuum could be traced back to an old stellar population.

One way to determine if the radio continuum emission is associated with star formation is to compare high resolution continuum images with molecular gas maps.  Since molecular gas is the raw material for star formation, it constrains the location and the amount of star formation in the galaxy.  If the radio continuum emission is extended and coincident with the molecular gas, that is strong evidence for star formation activity.  On the other hand, if the continuum emission is well described by a point source, whereas the CO is extended, then the radio continuum is more likely associated with an AGN or perhaps a nuclear starburst than with normal disk-type star formation.  If the radio emission is extended but in a very different location from the molecular gas, the situation is similar: the radio continuum is more likely to be powered by an AGN or the evolved stellar population. Up until now there has been very little knowledge of the CO morphology in early-type galaxies.  Recently, studies by Young (2000, 2001, 2002, 2005) have produced the largest sample of high resolution molecular gas maps of early-type galaxies to date.  

In this paper we present a high resolution search for radio continuum emission from Young's (2002) seven galaxy sample plus two dwarf early-types.  The radio maps are then compared to high resolution CO maps.  This information is then combined with radio morphology and FIR-radio correlation placement to trace recent star formation activity.  We also evaluate the Toomre Q parameter in order to address the question of gravitational stability. This project allows us the opportunity to independently corroborate previous optical and radio studies that suggest star formation activity in early-type galaxies and to probe the origin of the radio-FIR correlation in those galaxies.     

The layout of the paper is as follows: A sample description is given in section 2.  Existing and new observations are discussed in section 3 and 4, respectively.  The results of the new observations are given in section 5.  Radio continuum emission mechanisms are discussed in section 6.  The origin of FIR emission is discussed in section 7.  In sections 9 and 10, we discuss the implications of the results.  In section 11 we present the paper conclusions. 
\section{The Sample: FIR Bright Galaxies}
The IRAS satellite detected many early-type galaxies at 60 $\mu$m and 100 $\mu$m (Knapp et al. 1989).  The brightest ones (100 $\mu$m flux $>$ 1 Jy) have been searched for CO using single dish telescopes (Wiklind, Combes, and Henkel 1995 (Hence forth WCH95); Lees et al. 1991; Knapp \& Rupen 1996).  Young (2002) recently completed a high resolution survey of the molecular gas from seven bright CO detections of WCH95, using the Berkeley Illinois Maryland Association (BIMA) and the Owens Valley Radio Observatory (OVRO) interferometers.  Of the seven early-types observed, five were detected and resolved in CO.  The molecular gas in these five galaxies is located in very regular, symmetric rotating disk with diameters ranging from 2 to 12 kpc.  We chose the same seven-galaxy sample in which to image the radio continuum. We removed NGC 4649 from Young's (2002) original sample since it has recently been determined that the FIR from this galaxy is actually associated with its spiral companion NGC 4647 (Temi \etal~2004), and also no CO was detected by Young (2002).  Two dwarf elliptical companions of M31, NGC 205 and NGC 185, are added to the sample in order to explore the possibility of star formation in very low mass early-type galaxies.  These two galaxies have been imaged and detected in CO using the BIMA millimeter interferometer (Young 2000, 2001).  

All of the galaxies in our final sample have an r$^{1/4}$ profile or  ``a consistent classification as E or E/SO in several catalogs'' (WCH95).  In some cases, these classifications are based on photographic evidence.  Recently, CCD imaging suggests that one of our sample galaxies may have a more complicated morphology (NGC 5666, Donzelli \& Davoust 2003) so we prefer to refer to the galaxies in our sample as ``early-type'' galaxies.  The sample galaxies have a variety of properties and are found in a variety of environments.  Some of the galaxies are isolated; some are in loose groups, and some are deep within the Virgo Cluster.  Some have very smooth, regular morphologies, whereas others are classified as ``peculiar'' because of dust lanes, stellar shells and ripples, etc.  The sample galaxies have optical luminosities that range from 10$^8$ to 10$^{11}$ solar luminosities.  Table 1 contains ephemeris information for our galaxy sample.  
\section{Archival Radio Continuum Data}
VLA\footnote{The VLA is operated by the National Radio Astronomy Observatory, which is a facility of the National Science Foundation (NSF), operated under cooperative agreement by Associated Universities.} data with matching resolution and signal to noise at 20 cm to that of Young's (2000, 2001, 2002) CO maps exist for only three of the sample galaxies (NGC 3656, NGC 7468, and NGC 4476).  NGC 4476 suffers from confusion in the NVSS but is located within a deep multiconfiguration (D, C, and B) VLA map of M87 at 20 centimeters (F. Owen, priv. comm. 2001).  
\section{New 20cm VLA Observations}
We obtained new VLA 1.4 GHz C configuration data for the galaxies that are not detected in the NVSS.  For the purposes of spatial comparisons with Young's (2002) CO maps, new 20 cm observations were also obtained for NGC 807, NGC 5666, and UGC 1503 using the VLA B array.  We also re-observed NGC 3656 using the VLA B array in order to produce a map with similar noise level to the observations just mentioned.  Both the C and B array observations were made using two IF pairs, each consisting of 50 MHz each in left and right circular polarizations centered on two different frequencies.  This gives a total bandwidth of 200 MHz .  The requested rms noise for these observations is 0.01 mJy beam$^{-1}$.  The main parameters of the observed fields are listed in Table 3.  

The data were edited, calibrated, and imaged using standard tasks in the NRAO astronomical image processing system (AIPS; see van Moorsel, Kemball, \& Greisen 1996 for an up-to-date description). In the case of NGC 807 it was necessary to combine the C and B array data to minimize confusion due to the bright sidelobes of a nearby bright source.  The integrated radio fluxes, position angles, major axes, and minor axes for our detected sources (see table 4) were determined using the MIRIAD software package (Sault, Teuben, \& Wright 1995). Position angles and deconvolved source sizes were determined by fitting elliptical gaussians to the radio emission.  Integrated fluxes were summed in polygonal regions and in concentric ellipses.  Great care was taken to insure that no confusing sources contributed to the integrated flux.  
\section{Observational Results}
\subsection{Detections and Spatial Correlations}
A comparison of the CO and radio is discussed for each galaxy below:

\textbf{NGC 7468}- NGC 7468 is a member of a small group of galaxies.  It has been characterized as a very peculiar galaxy because of the presence of elliptical isophotes surrounding a nucleus that has been resolved into at least two sources of emission aligned along the major axis of the galaxy (Deeg, Duric, \& Brinks 1997).  Another extended optical feature connected with this galaxy is visible approximately 30\arcsec~ to the southwest (Nordgren \etal~1995).  Whitmore \etal~(1990) consider this galaxy to be a possible polar-ring galaxy, with these structures being part of the polar ring.  Recent broad band optical and H$\alpha$ images reveal the presence of several HII regions coincident with the two central optical components as well as H$\alpha$ emission associated with the polar ring (Cair$\acute{o}$s \etal~2001; Paz, Madore, \& Pevunova 2003).  Duprie \& Schneider (1996) detect a large HI disk (D$\sim$ 11$\arcmin$) using the 300 meter Arecibo Telescope.  NGC 7468 has been observed in CO at several single dish telescopes with varying results.  This galaxy has not been detected in the 2-1 line of CO (Lees \etal~1991), but there are several tentative detections from the IRAM 30 meter telescope in the 1-0 line (WCH95; Sage, Welch \& Young 2006).  Interferometric observations at 3 mm produced a non-detection (Young 2002).  However, an upper limit (derived from summing over the same spatial scale and velocity area as that of the single dish results) produces a flux consistent with that measured in the most recent single dish observations.  This is highly suggestive that the single dish detections may in fact be real, and that there is a high probability that NGC 7468 contains some molecular gas.

NGC 7468 already has sufficiently high resolution data in the VLA archive (Cox \& Sparke 2004).  In both sets of archive data the image quality suffers due to confusion from a bright source to the north.  The poor image quality is more pronounced in the 20 cm C array data and is perhaps due to its' poorer UV coverage.  The emission in this image is elongated in the same manner as the beam.  Both images show that the radio emission in this galaxy is resolved into two large regions (see Figure 1).  The northern blob is much larger, and contains more than 80$\%$ of the total flux.  The radio emission is coincident with the two central features observed in the aforementioned optical and H$\alpha$ images.  No radio continuum is detected along the polar ring.  Both sets of data show some low level radio emission that appears to extend outside the eastern part of the galaxy. The image quality is too poor to say whether this emission is real or just noise.  Perhaps deeper VLA observations can alleviate this difficulty in the future.

\textbf{NGC 3656}- This is a field early-type that shows indications of a gravitational interaction sometime in the near past (Balcells \etal~2001).  There are several small spherical objects, presumed to be dwarf galaxies, surrounding NGC 3656 (WCH95).  Young (2002) favors the explanation that the molecular gas was acquired during a merger event.  Both the CO and radio emission peak at the optical center, are elongated in the same position angle and have similar minor axis widths.  However, the radio emission is less extended in radius (by about 10\arcsec) than the CO in the northern parts of the galaxy.  The radio continuum also does not follow the CO extension curving off the south east side of the molecular disk.  The extension probably traces gas that has not yet reached dynamical equilibrium and so has not settled into the disk; we suspect that if the gas is not yet settled, star formation is unlikely to occur there.  A comparison of the molecular gas and the radio continuum can be found in Fig 2.

\textbf{NGC 5666}- This is an isolated, low-luminosity early-type galaxy.  This galaxy is known to have an extended HI disk (Lake, Schommer, \& van Gorkom 1987), and extended diffuse radio emission (Wrobel \& Heeschen 1988).  The radio continuum in NGC 5666 is extended on scales of a few kpc, and matches the CO extent along both axes rather well.  The CO is centrally peaked whereas the continuum intensity peak ($\sim$2 mJy beam$^{-1}$) is not located at the center, but to the southeast. A slice through the major and minor axes of the radio emission shows that the radio emission drops to $\sim$1.5 mJy beam$^{-1}$ inside a roughly circular region, $\sim$ 7\arcsec~ in diameter, centered on the optical nucleus.  The CO and radio continuum data are at matched resolution and so the absence of a central hole in the CO emission suggests that the underlying emitting structures are different, and so the mismatch at the center is not the result of a resolution effect.  A comparison of the molecular gas and the radio continuum can be found in Fig 3.  

A recent $B$-band image of NGC 5666 from Donzelli \& Davoust (2003) shows a ring of star formation, 10\arcsec\ in diameter and coincident with the ring of radio continuum emission in Figure 3.  This irregular ring contains some fragments which are suggestive of being spiral arm segments, all within the region where CO emission is found.  The galaxy's spectrum is dominated by an old stellar population, however (Donzelli \& Davoust 2003).  Thus it appears that some flocculent spiral structure may be developing in the center of this gas rich early type galaxy.

\textbf{UGC 1503}- This is a an isolated early-type galaxy.  The radio emission in this galaxy is extended on scales of $\sim$5 kpc, and is centered on the optical center of the galaxy.  All of the emission is contained in a ring with a central circular depression ($\sim$5$\arcsec$~in diameter) where the radio emission nearly drops to zero.  The peak of the radio emission ($\sim$ 0.3 mJy beam$^{-1}$) occurs in several places along the ring whereas the CO emission is centrally peaked.  The extent of the CO and radio emission are roughly coincident along both the galaxy's major and minor axes. A comparison of the molecular gas and the radio continuum can be found in Fig 4.

\textbf{NGC 807}- This is an isolated giant early-type galaxy. The CO distribution is asymmetric with most of the molecular gas ($\sim$ 70$\%$) located in
the southeast section of the galaxy. Young (2002) suggests that shear
from the differential rotation in the disk would destroy the asymmetry
on a time scale of 10$^9$ years.  Since the asymmetry is still
present, this may be an indication that the gas was acquired in an
interaction less than 10$^9$ years ago.  The radio continuum in NGC
807 is extended on scales of ~9 kpc, and matches the CO extent along
both axes rather well. There is an asymmetry in the distribution of
the continuum much like that of the CO distribution (most of the
emission, $\sim$ 60$\%$, is located in the southeast).  In fact the
radio emission ($\sim$0.3 mJy beam$^{-1}$) peaks at roughly the same
location as the CO emission.  However, it may be possible that the
perceived asymmetry in the CO is in reality due to differing
CO-to-H$_{2}$ conversion factors existing on alternate sides of the
galaxy.  If this is the case, then the correlation of the two emission
peaks may be nothing more than a coincidence.  However, the high
degree of morphological similarity between the CO and radio continuum present in several of the other sample galaxies suggest it is more than a coincidence.  A comparison of the molecular gas and the radio continuum can be found in Fig 5.
 
\textbf{NGC 185}-  This is a dwarf elliptical companion of M31.  The molecular gas in this galaxy is found in large complexes similar to those observed in the Milky way (Young 2000, 2001).  An HI and optical study of the interstellar medium (ISM) by Young \& Lo (1996) shows that complex dust regions near the center of the galaxy are coincident with the detected CO emission, and the HI emission exhibits a very clumpy distribution.  We detect a small amount of radio emission (peak intensity is 5$\sigma$) in our image of NGC 185.  The small patch of emission is slightly offset from the center of the galaxy and has a flux of $300 \pm 100~\mu$Jy (see Figure 6).  The emission is roughly coincident with a region of extended H$_\alpha$ emission detected by Young \& Lo (1997).  The surface brightness, luminosity, diameter, and morphology of the H$\alpha$ emission are within the ranges of known supernova remnants (e.g., Long \etal~1990). Supernova remnants in M31 have 20 cm radio fluxes that range from S$_\nu$=150-2600 $\mu$Jy (Braun \& Walterbos 1993).  Since M31 and NGC 185 are at relatively the same distance we conclude that the small patch of radio emission probably originates from a supernova remnant inside the galaxy.  The small spatial offsets between the CO detections and the radio continuum emission would not be noticeable in the other sample galaxies at 18--60 Mpc.
\subsection{Nondetections}
\textbf{NGC 4476}-  This galaxy is a small early-type 12$\arcmin$~ in projection from M87 near the center of the Virgo Cluster.  The stellar kinematics are normal for an elliptical galaxy with a central velocity dispersion of 132~km~s$^{-1}$. Dust is clearly seen on optical photographs, showing the presence of a nuclear dust-ring (Tomita \etal~2000).  Young (2002) favors an external origin for the detected CO emission based on the fact that the specific angular momentum of the gas is three times the specific angular momentum of stars.  VLA 20 cm continuum observations of NGC 4476 in the B, C, and D configurations yield an upper limit of 5$\times$10$^{-30}$ W m$^{-2}$ Hz$^{-1}$ or 0.5 mJy (F. Owen, priv. comm. 2001).  No previous radio detections have been reported for this galaxy.

\textbf{NGC 205}-  This galaxy is another dwarf elliptical companion to M31.  Similarly to NGC 185, the molecular gas in this galaxy is found in large complexes similar to those observed in the Milky way (Young 2000, 2001).  An HI and optical study of the interstellar medium (ISM) by Young \& Lo (1996) shows that complex dust regions near the center of the galaxies are coincident with the detected CO emission, and the HI emission exhibits a very clumpy distribution.  No radio continuum emission was detected in the image for NGC 205.  An upper limit to the radio continuum brightness in this galaxy is determined from a 3 sigma upper limit of the rms noise ($\sim$ 0.02 mJy beam$^{-1}$).  The flux for NGC 205 is located in Table 4.  No previous radio detections have been reported for this galaxy.

\section{The Origin of Radio Continuum Emission}
There are two major components of the radio continuum emission in galaxies.  These are synchrotron emission (non-thermal) and free-free emission (thermal bremsstrahlung).  In star-forming galaxies, synchrotron emission comes from cosmic ray (CR) electrons spiraling in the magnetic field of a galaxy.  Type II and Type Ib supernovae and their remnants are thought to be the main electron accelerators (Kennicutt 1989, Condon 1992).  Often, the synchrotron spectra can be approximated by a power law in some limited frequency range.  When this occurs, the spectral index $\alpha$ is defined as S$\propto\nu^{-\alpha}$, where S is the integrated flux density and $\nu$ is the frequency. 

Thermal emission originates from shocked regions around supermassive stars where superheated electrons collide with the ISM.  Usually the optically thin thermal emission dominates over the optically thick.  Optically thin thermal radiation has a characteristic spectrum ($\alpha\leq\sim$0.5), and so it should be distinguishable from the steeper-spectrum non-thermal emission.  Typically $\alpha$ is determined by comparing the total flux densities obtained at two frequencies with closely matching resolutions.  For example:
\begin{equation}
\log\left(\frac{S_{1.4_{GHz}}}{S_{4.89_{GHz}}}\right)=-\alpha\log\left(\frac{\nu_{1.4_{GHz}}}{\nu_{4.89_{GHz}}}\right)
\end{equation}
The available data allow us to calculate the spectral index for three of our sample galaxies: NGC 5666, NGC 7468, and NGC 3656 (see Table 4).  The spectral indices determined from the total fluxes in the two frequency bands for these three galaxies range from 0.5 to 0.8.  Pixel by pixel maps of the spectral index do not show any evidence of a central point source of significantly different spectral index.  We note here that in order to avoid variability effects, this type of comparison should be done using observations made with identical visibility coverage on the same day.  We have not done this, so the calculated spectral indices are only rough estimates.    

The high resolution radio images rule out nuclear monsters both spectroscopically ($\alpha\sim$~1 for luminous AGN; Zhang, Lin, \& Fan 2003) and morphologically (Only NGC 3656 exhibits a centrally peaked radio distribution).  It is possible that our galaxies contain very low radio power AGN.  We estimate the limit to the flux that an unresolved AGN would contribute to the total radio power by measuring the radio flux contained inside one beam centered on the optical centers of our galaxies.  We find that these limits are only $\sim$10 percent of the total flux and so any contribution from a central point source to the total radio power is small. 

\section{The Origin of Far Infrared Emission}
\subsection{The Radio-Far Infrared (FIR) Correlation}
The correlation between the FIR luminosity and the radio continuum emission observed for normal spiral galaxies is extremely tight and roughly linear ($<$q$>=$2.3 at 1.4 GHz; Helou, Soifer, $\&$ Rowan-Robinson 1985) which implies that the process of star formation produces both types of emission.  It is assumed that the FIR and radio emission comes from dust heated by massive stars and CR from supernovae remnants of the same massive stars, respectively.

The FIR-radio correlation has been observed to be non-linear at low frequencies where nonthermal emission dominates ($\nu<$5 GHz) as well as for optically-selected samples containing low-luminosity ($\leq~\times10^9 L_\odot$) galaxies (Fitt, Alexander, \& Cox 1988, Devereux $\&$ Eales 1989, Condon 1992). One explanation for this is that low-luminosity galaxies are less massive than normal galaxies, and are more likely to lose their CR to convection (Chi \& Wolfendale 1990). Subsequently, a small galaxy would appear under-luminous in the radio with respect to the FIR emission, and the correlation would become non-linear even if the low luminosity galaxies are forming stars.  It has also been shown that galaxies with low L$_{60\mu~m}$ also drop off the FIR-radio correlation (Yun Reddy \& Condon 2001). 

Deviations from the correlation may also be due to changes in the surrounding environment such as the presence of an intercluster medium (ICM).  The magnetic fields in galaxies can become compressed as they move through a dense ICM.  This compression may significantly alter the amount of observed synchrotron emission (Reddy \& Yun 2003) and should enhance the radio continuum emission in the galaxy, making it appear over-luminous in the radio.

The FIR flux can be approximated from the IRAS 100 and 60 micron flux densities using the following relation:
\begin{equation}
\left(\frac{FIR}{W m^{-2}}\right)=1.26\times10^{-14}\left(\frac{2.58S_{60 \mu m}+S_{100\mu m}}{Jy}\right)
\end{equation}
where S$_{100 \mu m}$ and S$_{60 \mu m}$ are the 100$\mu$ m and 60$\mu$ m FIR fluxes in Janskys (Condon 1992). Helou, Soifer, \& Rowan-Robinson (1985) defined the ratio between the far-infrared and radio continuum emission as follows:  
\begin{equation}
q=log\left(\frac{FIR}{3.75\times10^{12}W m^{-2}}\right)-log\left(\frac{S_\nu}{W m^{-2}Hz^{-1}}\right)
\end{equation}
for $\nu$=1.4 GHz.  

An evaluation of the ratio q for the galaxies in our sample show that five (NGC 3656, NGC 5666, NGC 7468, NGC 807, and UGC1503) of the eight galaxies sit within 3$\sigma$ of the radio-FIR correlation defined for star forming spirals.  The median q value for the sample galaxies detected in radio continuum is $<$q$>=$2.7$\pm$0.1 (see Figure 7).  We would also like to note that our sample galaxies with the lowest optical luminosities ($<~10^9 L_\odot$) also have the lowest FIR luminosities.  

Condon, Cotton, and Broderick (2002) find that nearly all galaxies with spectroscopically or morphologically determined AGNs have a FIR spectral index $\alpha_{FIR}$ $<$ 1.5.  The FIR spectral index can be determined as follows:
\begin{equation}
\alpha_{FIR}=\log\left(\frac{S_{60\mu m}}{S_{25\mu m}}\right)-\log\left(\frac{60}{25}\right)
\end{equation}
The galaxies in our sample have FIR spectral indices greater than 1.5 (see Table 5), which is another indication that these galaxies do not contain AGN or at least the FIR emission does not have a significant contribution from a dust enshrouded AGN.  This result is consistent with the radio morphologies of the sample galaxies, which also indicates that luminous AGN are not present in these galaxies.
\section{Star Formation Rates and Efficiencies}
The star formation rate (SFR) is calculated using the method derived by Condon, Cotton, and Broderick (2002) for spiral galaxies.  The FIR luminosity is proportional to the SFR as:
\begin{equation}
\left(\frac{L_{FIR}}{L_\odot}\right)\sim~1.1\times10^{10} \left[\frac{SFR (M \geq 5 M_\odot)}{M_\odot~yr^{-1}}\right]
\end{equation}
This is the star formation rate for stars of M $\geq$ 5 M$_\odot$.  To calculate the star formation rate for all the stars between 0.1 to 100 M$_\odot$ we simply multiply this result by a factor of 5.5 (for a Salpeter initial mass function).  The average star formation rate for stars between 0.1 and 100 M$_\odot$ for the sample galaxies detected in radio continuum is $\sim$ 2 M$_\odot$ year$^{-1}$ (see Table 5).

The gas depletion time scale can be determined by dividing the H$_2$ masses found by Young (2002) by the star formation rate.  The H$_2$ masses for the sample galaxies can be found in Table 6.  The average depletion time scale for our sample galaxies detected in CO is $\sim$~0.7 Gyr (see Table 5) which is much smaller compared to typical gas depletion times scales found in spiral galaxies ($\sim$~2 Gyr; Kennicutt 1983).  The smaller depletion timescales may indicate that star formation in early-type galaxies is more efficient than in normal spiral galaxies.  There are several other possibilities for the smaller depletion time scales.  The star formation could be episodic with prolonged periods between star formation events either due to differing ISM conditions throughout the galaxies or to the fact that some molecular gas is still in falling from outside of the galaxies (Pipino \etal~2005; Ferreras \& Silk 2003; and many others) or both. It is also possible that the gas masses are incorrectly estimated (This will be discussed further in section 10).

\section{Discussion}
\subsection{Candidates for Recent Star Formation: Galaxies Consistent with the radio-FIR Correlation}
The five best candidates for star formation in our sample are NGC 3656, NGC 5666, NGC 807, NGC 7468, and UGC 1503.  These galaxies sit within 3$\sigma_q$ of the radio-FIR correlation defined for star forming spirals.  NGC 3656, NGC 5666, NGC 807, and UGC 1503 have CO emission that is roughly coincident with the observed radio continuum emission.  In fact, the distribution of the continuum emission in these galaxies resembles those of normal spiral galaxies, and probably traces a ring of star formation.   NGC 3656 has a star formation rate a factor of two greater than the average for our sample.  We might expect this since NGC 3656 has undergone some sort of interaction in the past, which may have produced tidally-induced star formation.

Optical and H$\alpha$ images show that NGC 7468 possess star forming regions much like those in our own galaxy (Cair$\acute{o}$s \etal~2001).  The radio emission is coincident with two large patches of H$_{\alpha}$ emission.  NGC 7468 sits right on the radio-FIR correlation.  However, the star formation rate for NGC 7468 (0.8 M$_\odot$ year$^{-1}$) is approximately a factor of three less than the average of our sample.  This smaller SFR could indicate that most of the molecular gas has been used up in a recent star formation event, and the SFR has slowed down to its present value.  The lower star formation rate is also consistent with the fact that this galaxy contains two orders of magnitude less molecular gas than the other galaxies in the sample. 

Our results indicate that star formation is occurring in more than half of our sample galaxies. We note here that our galaxy sample is FIR and CO selected, and so it may not be representative of early-type galaxies.  However, if we consider all of the early-type galaxies in the sample of Cotton, Condon \& Broderick (2002) we find that many of them do lie on the radio-FIR correlation (see Figure 7).  Their galaxies are both radio and FIR selected, not CO selected.  Full detections that sit within 3$\sigma$ of $<$q$>$$=$2.3 include $\sim$15 galaxies, suggesting that our sample galaxies may not be that unusual.
\subsection{Deviant Galaxies}  
Interferometric images of the molecular gas in NGC 185 and NGC 205 find objects much like the giant molecular clouds in our own galaxy, in terms of sizes and masses (Young 2000, 2001).  Strangely, these galaxies show significant deviations from the spiral galaxy radio-FIR correlation in the sense that they are radio faint (they are the two symbols at the bottom left of Fig 7).  The radio nondetection found for NGC 205 and the small amount of radio emission detected in NGC 185 are consistent with studies of low surface brightness dwarfs (Hoeppe et al. 1994).  NGC 205 and NGC 185 most likely deviate from the radio-FIR correlation because they do not posses enough mass to retain all of their CR.  The superbubbles created by supernovae are able to expand out of the galaxy, compressing the magnetic field.  It is easier for the CR to escape if the magnetic fields are located in shocks produced by the expanding bubbles (Chi $\&$ Wolfendale 1990).  Also, the SFR may be slow enough so that the CR from defunct SNR have already long since escaped the galaxy (Klein \etal~1992).  

NGC 4476 is the most puzzling galaxy in our sample.  There is certainly enough molecular gas in this galaxy to form stars.  Why then is there no detected radio emission?  The star formation rate in this galaxy is low, but $L_{FIR}/L_B$ is similar to the star-forming NGC 807.  One explanation for the lack of radio continuum may be that most of the atomic hydrogen has been removed due to ram pressure as it travels through the ICM surrounding M87 (Lucero, Young, \& van Gorkom 2005).  Perhaps the bulk of the magnetic field has been removed along with the HI in the stripping event, and consequentially the radio is not observed.  A similar phenomenon appears to to be occuring in the Virgo Cluster spiral galaxy NGC 4522.  This galaxy has a large amount of extraplanar HI, and radio continuum is indeed associated with the extraplanar gas (Vollmer \etal~2007).

\subsection{Dynamical Mass and Gravitational State of the Gas Disks}
Okuda \etal~(2005) recently have shown that the giant radio galaxy 3C~31 (NGC 383) contains a massive molecular gas disk, though little star formation activity has been found (Owen, O'Dea, \& Keel 1990).  Okuda \etal (2005) argue that star formation is not occurring because the galaxy's deep potential stabilizes the gas disk against collapse, an idea which was also put forward by Kennicutt (1989) for early type galaxies in general.  The stability of a rotating gas disk can be assessed quantitatively by Toomre's criterion as revised by Sakamoto \etal~(1999),
\begin{equation} 
Q=\frac{\kappa\sigma}{\Omega V}\left(\frac{M_{gas}}{M_{dyn}}\right)^{-1}
\end{equation}
where
\begin{equation}
M_{dyn}=\frac{RV^2}{G};~M_{gas}=\Sigma_{gas}\pi R^2
\end{equation}
and where $\sigma$ is the velocity dispersion, G is the gravitational constant, $\Sigma_{gas}$ is the gas surface density, $\Omega$ is the angular velocity, V is the gas velocity, and $\kappa$ is the epicyclic frequency.  Writing the $Q$ parameter in this way elucidates the role of the galaxy's dynamical mass and the gas to dynamical mass ratio;  values of $Q > 1$ indicate gravitational stability.  

For the estimation of $Q$ we assume our galaxies have an axisymmetric potential in which the velocity of the gas either rises linearly ($\kappa=2\Omega$) or is constant ($\kappa=\sqrt{2}\Omega$) out to the edge of the gas disk (Binney \& Tremaine 1987, pp. 120-122).  Gas and dynamical masses are taken from Young (2002) and can be found in Table 6.  Molecular hydrogen masses are multiplied up by a factor of 1.36 to account for the presence of helium.  The true velocity dispersion of the gas is not known so a typical value of 10 km s$^{-1}$ is assumed.  The ratios M$_{gas}$/M$_{dyn}$ for this sample range from 0.02 to 0.1, similar to the values found by Koda \etal\ (2005) in early type spirals.  The values of the $Q$ parameter range from 0.8 to 4 under these assumptions.

Only one of the estimated $Q$ values is less than 1 (indicating instability), but the remainder are sufficiently close to 1 that they may also be consistent with gravitationally unstable disks.  The simplified analysis described above neglects the destabilizing influence of the stars in the galaxy.  Jog \& Solomon (1984) show that the interactions between gas and stars can make the two-fluid system gravitationally unstable even if the gas is stable when considered by itself (as is done above).  Thus the values of $Q$ which are somewhat higher than 1 should not necessarily be interpreted as stability.  We must also note that local surface densities are undoubtedly higher than average at some points in the galaxies and lower at others.  A more detailed analysis is precluded by the fact that $Q$ scales with the velocity dispersion, which might be a factor of two different from what we have assumed (e.g. Okuda \etal\ 2005; Jog \& Solomon 1984).

We have shown that the existence of a Toomre-type gravitational instability in these molecular disks cannot be ruled out, and the moderate to low $Q$ values found here may be consistent with the other evidence suggesting star formation activity in these galaxies.  In particular, NGC 3656 has the lowest $Q$ value in the sample, and it has one of the highest star formation rates and $L_{FIR}/L_B$ values.

\subsection{Timescales}
Recently it has been suggested that that the star formation process itself may have an effect on a galaxies position on the radio-FIR correlation.  Several studies comparing the spatial extent of radio continuum and FIR emission in spiral galaxies suggest that the diffusion scale length of CR electrons is much larger than the mean free path of the dust heating photons (Bicay \& Helou 1990; Murphy \etal~2006).  This means that the continuum emission will become more and more smoothed out compared to the FIR emission as the CR diffusion timescales increase. These studies also show that the CR diffusion timescales tend to increase as the disk averaged SFR in the galaxy decreases (Murphy \etal~2006).  So if a galaxy has a SFR below some threshold value, its radio continuum emission could become so smoothed out that it may be hard to detect.  This effect would be amplified if a galaxy is not massive enough to retain its CR, has had its magnetic field removed, or experiences episodic star formation.  It is interesting to note that the galaxies in our sample with the smallest SFR ($\leq$~0.2 M$_{odot}$ yr$^{-1}$) also appear to contain little to no radio continuum.  It may be that these galaxies deviate from the radio-FIR correlation simply because their SFR are too low.

\subsection{The Origin and Fate of the Molecular Gas}
If the star formation proceeds at rates inferred from the far infrared emission, the transformation from gas to stellar disks will occur over a few billion years.  According to Young (2002), these disks will be rotationally supported, on the order of a few kpc, will contain ~1$\%$ of the total stellar mass, and will be very similar to the stellar disks known to exist in many early-types (Scorza \& Bender 1995; Emsellem \etal~2004).  Our results clearly show that stellar disks can grow out of gas disks in at least some cases.

The gas in these disks could be produced by either internal processes (Temi \etal~2007) or it could be left over from mergers (Barnes 2002; Naab, Jesseit \& Burkert 2006). In other words, the presence of a star forming gas disks in spheroidal galaxies is consistent with both the early-assembly (Monolithic collapse) and the late assembly (hierarchical growth) pictures.  While our results can neither confirm of rule out either scenario they do emphasize that the morphology of an early-type galaxy, parameterized by its bulge-to-disk ratio, may evolve somewhat through the process of star formation even while the galaxy remains on the red sequence.  It would be very interesting to see if stellar disks are detectable in early-type galaxies at high redshift.

\section{A Cautionary Note}
The H$_2$ mass in galaxies is estimated using CO(1-0) line emission.  The standard conversion factor used for molecular clouds in the Milky way is given by N(H$_2$)/I$_{CO}$ = 3$\times$10$^{20}$ cm$^{-2}$ K km s$^{-1}$)$^{-1}$ (Sanders, Solomon, $\&$ Scoville 1984).  The use of this conversion factor assumes that the mean properties of the molecular gas in the early-type galaxies in the sample are similar to those in the Milky Way.  The molecular clouds in early-types might not exhibit the same properties, such as temperature and density, as those found in the Milky Way (Georgakakis \etal~2001).  It is expected that such differences will modify the CO to H$_2$ conversion factor.  If the molecular gas densities in early-types are larger than in the Milky way, the conversion will underestimate the the H$_2$ mass.  If instead the densities are the same, but the temperatures are lower than those in the Milky way, the conversion factor will overestimate the H$_2$ mass.  

A modified H$_2$ mass will affect the depletion timescales as well as the gas mass to dynamical mass ratios.  If gas masses are overestimated then the depletion timescales become even faster, and the gravitational stability greater.  In order to achieve depletion time scales like those in spiral galaxies the gas mass would have to be underestimated by a factor of 4.
\section{Conclusions}
In this paper we present new high resolution radio continuum observations of eight FIR and CO selected early-type galaxies, and compare them to high resolution CO maps in order to investigate the importance of star formation in these galaxies.  In summary:\\\\
$\bullet$ Five of the eight galaxies (NGC 807, NGC 3656, NGC 5666, NGC 7468 and UGC 1503) sit within 3$\sigma_q$ of  the radio-FIR correlation defined for star forming spirals.\\\\
$\bullet$ The radio continuum is extended on scales of a few to $\sim$9 kpc.  Good spatial correlations between the CO and radio emission exist in four of the sample galaxies (NGC 807, NGC 5666, UGC 1503, and NGC 3656).  This morphological similarity confirms Wrobel \& Heeschen's (1988) suggestion that early-type galaxies on the FIR-radio correlation may be forming stars.  Many of the early-type galaxies in the NVSS catalogue also sit on the radio-FIR correlation.  These results are in agreement with other recent work suggesting that star forming early-type galaxies are more common than has been thought in the past.\\\\
$\bullet$ The average star formation rate for the galaxies detected in radio continuum is $\sim$ 2 M$_\odot$ year$^{-1}$.  The average gas depletion time scale for the galaxies detected in CO is $\sim$ 0.7 Gyr, which is much less than the gas depletion time scales observed in star forming spiral galaxies ($\sim$~2 Gyr).\\\\
$\bullet$ A simple gravitational stability estimate suggests that the molecular disks are likely to be unstable, driving the inferred star formation activity.\\\\
$\bullet$ The molecular gas in our star forming early-type sample galaxies will produce stellar disks containing $\sim$~1$\%$ of the total stellar population, and radii of a few kpc.\\\\
$\bullet$ A few of the sample galaxies (NGC 4476, NGC 185, and NGC 205) deviate significantly from the FIR-radio correlation in that they are radio faint.  In the case of the two dwarf early-types, their small masses may allow a significant number of cosmic rays to escape.  In the case of NGC 4476, the large scale magnetic field may have been removed along with HI gas due to the passage through a hot intercluster medium. 

\acknowledgments
We thank the anoymous referee for a very constructive report, and Joan Wrobel for her insightful comments which have made this work much stronger.  This research has made use of the NASA/IPAC Extragalactic Database (NED) which is operated by the Jet Propulsion Laboratory, California Institute of Technology, under contract to the national Aeronautics and Space Administration.  Last but not least, this work has been supported by NSF grants AST-0074709 and AST-0507432 to New Mexico Institute of Mining and Technology. Lisa M. Young thanks New Mexico Tech for a sabbatical leave and is grateful to the University of Oxford sub-department of Astrophysics for their hospitality.
\clearpage
\bibliographystyle{apj3}
{\footnotesize
     \bibliography{ms}
     }     
\clearpage
\begin{table}
\begin{scriptsize}
\caption{\bfseries Sample Galaxies}
\begin{tabular}{lcccccl}
\\
\tableline\tableline
Galaxy     &Type$^a$ &RA             &DEC          &L$_B$                 &Distance &Environment\\
           &         &               &             &(10$^9$~L$_{\odot}$)  &(Mpc)    &\\
\tableline 
NGC185     &dE3p     &00h38m57.5s    &48d20m12s    &0.057                 &0.63     &Group \\
NGC205     &E5p      &00h40m22.0s    &41d41m07s    &0.30                  &0.85     &Group \\
UGC1503    &E        &02h01m19.8s    &33d19m46s    &17                    &69       &field \\
NGC807     &E        &02h04m55.7s    &28d59m15s    &32                    &64       &field \\
NGC3656    &IO       &11h23m38.4s    &53d50m31s    &16                    &45       &merger remnant \\
NGC4476    &E/SO     &12h29m59.1s    &12d20m55s    &3.5                   &18       &Virgo cluster\\
NGC5666    &Ec       &14h33m09.2s    &10d30m39s    &6.3                   &35       &field \\
NGC7468    &E3p      &23h02m59.2s    &16s36m19s    &4.0                   &28       &Group \\
\tableline
\end{tabular}
\begin{tabular}{l}
*Environment descriptions, distances, and blue luminosities are\\
taken from Wiklind, Combes, $\&$ Henkel 1995.  For the case of\\
NGC 185 and NGC 205, the environment descriptions and distances \\
are the same as used in Young (2000, 2001).  The blue luminosities\\
from WCH95 have been corrected for these distances.\\ 
$^a$Galaxy classifications are taken from from Knapp \etal~1989\\
\end{tabular}
\end{scriptsize}
\end{table}
\clearpage
\begin{table}
\begin{scriptsize}
\caption{\bfseries Archive Radio Continuum Data.}
\begin{tabular}{lccccccccc}
\\
\tableline\tableline
Galaxy &Config &Observation   &Flux        &TOS$^a$   &$\nu$       &Program  &Noise$^b$              &Beam           &Flux$^c$\\
       &       &Dates         &Calibrator  &(min) &(10$^9$ Hz) &ID      &(mJy beam$^{-1}$)  &(\arcsec)      &(mJy)\\
\tableline
\\
N185     &D  &1994-1996              &3C295          &0.5    &1.4         &NVSS    &0.5        &45$\times$45        &$<$ 1.5\\
 & & & & & & & & &$-$\\
N205     &D  &1994-1996              &3C295          &0.5    &1.4         &NVSS    &0.5        &45$\times$45       &$<$ 1.5\\
 & & & & & & & & &$-$\\
U1503    &D  &1994-1996  &3C295  &0.5    &1.4  &NVSS    &0.5   &45$\times$45    &$<$ 1.5\\
 & & & & & & & & &$-$\\
N807     &D  &1994-1996              &3C295          &0.5    &1.4         &NVSS    &0.5        &45$\times$45       &$<$ 1.5\\
 & & & & & & & & &$-$\\
N3656    &D  &1994-1996  &3C295  &0.5    &1.4  &NVSS    &0.5   &45$\times$45    &20.7\\
 & & & & & & & & &1.2\\
                 &C  &1987 Feb   &3C286  &24     &1.46 &$^1$AM197   &0.5   &71$\times$27 &20.8\\
 & & & & & & & & &1.6\\
             &C  &1990 Nov   &0134+329 &59   &4.89 &$^2$AZ45    &0.03  &4.3$\times$3.6 &8.8\\
 & & & & & & & & &0.5\\
         &B  &1997       &3C286    &3    &1.44 &FIRST   &0.2   &5.4$\times$5.4 &23.2\\
 & & & & & & & & &1.0\\
N4476    &B &2000           &3C286          &2321       &1.4         &$^5$AO136/144/149   &0.2      &4.5$\times$4.5   &$<$ 0.5\\
 & & & & & & & & &$-$\\
N5666    &D  &1994-1996              &3C295          &0.5    &1.4         &NVSS    &0.5        &45$\times$45       &20.6\\
 & & & & & & & & &1.5\\
           &C  &1985 Sep      &1328+307       &11     &4.89        &$^4$AW137   &0.09       &5.6$\times$4.7   &6.0\\
 & & & & & & & & &0.7\\
N7468    &D  &1994-1996              &3C295          &0.5    &1.4         &NVSS    &0.5       &45$\times$45       &11.2\\
 & & & & & & & & &2.2\\
 &C  &1994 Oct      &0134+329       &28     &4.89        &$^3$AC396   &0.06      &4.1$\times$3.3   &4.5\\
 & & & & & & & & &0.5\\                
 &B  &1994  Aug      &1328+307    &23     &1.46       &$^3$AC394   &0.1      &7.2$\times$4.3     &8.6\\
 & & & & & & & & &0.7\\
\tableline
\end{tabular}
\begin{tabular}{l}
$^a$TOS is the total time on source galaxy in minutes.\\
$^b$The rms back ground noise in the individual maps in units of (mJy beam)$^{-1}$.\\
$^c$Flux upper limits are 3$\sigma$ of the rms noise in the image.\\
NVSS:~Condon \etal~1998\\
FIRST:~White \etal~1997\\
\hline
\bf{Table References:}\\
1)~Moellenhoff, Hummel, \& Bender 1992\\
2)~Gregorini, Messina, \& Vettolani 1990\\
3)~Cox \& Sparke 2004\\
4)~Wrobel \& Heeschen 1991\\
5)~Owen, priv. comm. 2001
\end{tabular}
\end{scriptsize}
\end{table}

\clearpage
\begin{table}
\begin{scriptsize}
\caption{\bfseries New 20cm Radio Continuum Observations.}
\begin{tabular}{lcccccccc}
\\
\tableline\tableline
Galaxy &Config &Observation   &Flux             &TOS$^a$    &Noise             &Beam        &Flux &$\sigma$\\
       &       &Dates         &Calibrator       &(min)  &(mJy beam$^{-1}$) &(\arcsec)   &(mJy) &(mJy)\\
\tableline
\\
NGC185   &C    &2001 Jul           &0137+331         &49    &0.03             &14.6$\times$12.0      &0.3 &0.1\\\\
NGC205   &C    &2001 Jul           &0137+331         &48    &0.02             &14.5$\times$12.4      &$<$ 0.06 &$-$ \\\\
UGC1503  &C    &2001 Jul           &0137+331         &52    &0.05             &14.2$\times$12.9      &2.5 &0.5\\\\
         &B    &2002 Aug           &1331+305         &64    &0.03             &5.0$\times$4.5        &2.3 &0.8\\\\
NGC807   &C    &2001 Jul           &0137+331         &41    &0.03             &15.8$\times$13.5      &0.86 &0.3\\\\
         &B    &2002 Aug           &1331+305         &73    &0.06             &4.7$\times$4.4        &$<$0.2  &$-$\\\\
NGC3656  &B    &2002 Jun           &1331+305         &46    &0.03             &4.5$\times$3.8        &18.4 &1.2\\\\
NGC5666  &B    &2002 Jun           &1331+305         &46    &0.04             &4.6$\times$4.4        &16.5 &0.8\\\\
\tableline
\end{tabular}
\begin{tabular}{l}
$^a$TOS is the total time on source galaxy in minutes.\\
\end{tabular}
\end{scriptsize}
\end{table}
\clearpage
\begin{table}
\begin{scriptsize}
\caption{\bfseries Highest Resolution Radio Morphology}

\begin{tabular}{lcccccccc}
\tableline\tableline 
           &\multicolumn{6}{c}{20cm Radio Observations}  &6cm Radio &\\
\tableline
           &20cm Flux &PA        &Maj Axis$^c$ &Min Axis$^c$ &Maj Axis &Min Axis  &6cm Flux         &\\
Galaxy     &$\sigma$  &($\circ$) &(\arcsec)    &(\arcsec)    &(kpc)    &(kpc)     &($\sigma$) &$\alpha^a$\\
           &(mJy)     &          &             &             &         &          &(mJy)    &($\sigma$)\\     
\tableline
$\star$N185    &0.3     &$-$   &$-$  &$-$   &$-$ &$-$  &$-$   &$-$\\
         &0.1     &        &       &       &     &      &      & \\
N205     &$<$ 0.06    &$-$     &$-$    &$-$    &$-$  &$-$   &$-$   &$-$\\
         &$-$     &        &       &       &     &      &      & \\
U1503    &2.3     &70.0    &23.39  &16.94  &7.8  &5.7   &$-$   &$-$\\
         &0.8     &        &       &       &     &      &      & \\
N807$^b$ &1.2     &$-$55.5 &28.83  &13.53  &8.9  &4.2   &$-$   &$-$ \\
         &0.4     &        &       &       &     &      &      & \\
N3656    &18.4    &$-$4.3  &6.06   &2.65   &1.3  &0.6   &8.8   &0.6\\
         &1.2     &        &       &       &     &      &0.5   &0.2\\
N4476    &$<$ 0.5 &$-$     &$-$    &$-$    &$-$  &$-$   &$-$   &$-$\\
         &$-$     &        &       &       &     &      &      & \\
N5666    &16.5    &$-$21.7 &12.52  &11.38  &2.1  &1.9   &6.0   &0.8\\
         &0.8     &        &       &       &     &      &0.7   &0.2\\
N7468    &8.6     &$-$63.4 &13.03  &11.44  &1.8  &1.6   &4.5   &0.5\\
         &0.7     &        &       &       &     &      &0.5   &0.3 \\
\tableline
\end{tabular}
\begin{tabular}{l}
$\star$The emission in NGC 185  is unresolved.\\
$^a\alpha$ is the spectral index, determined as $\log\left(\frac{S_{1.4_{GHz}}}{S_{4.89_{GHz}}}\right)=-\alpha\log\left(\frac{\nu_{1.4_{GHz}}}{\nu_{4.89_{GHz}}}\right)$\\
$^b$The flux and source sizes quoted here are for the combined B and C VLA archive data.\\
The beam size of the final image is 10.9$\times$8.8 with an rms noise of 0.05 mJy beam$^{-1}$\\
$^c$We note here that UGC 1503, NGC 807, NGC 5666, NGC 7468 are not well fitted by gaussians\\
and so the their true sizes may be underestimated.\\
\end{tabular}
\end{scriptsize}
\end{table}
\clearpage
\begin{table}
\begin{scriptsize}
\caption{\bfseries Properties of the Interstellar Medium}
\begin{tabular}{lccccccl}
\\
\tableline\tableline
Galaxy  &L$_{1.4GHz}$      &L$_{FIR}$           &q          &$^c\alpha_{FIR}$ &SFR &T$_D^a$               &L$_{FIR}$/L$_B$\\
        &(W Hz$^{-1}$)     &(10$^8$ L$_\odot$)  &$\sigma_q$ &                 &(M$_\odot$ yr$^{-1}$)      &(10$^8$years) &\\
\tableline
N185  &1.9$\times$10$^{16}$        &4.47$\times$10$^{-3}$      &3.4       &$>$3.4      &2.2$\times$10$^{-4}$   &2.5     &0.01\\
      &                            &           &0.3       &            &                       &        &\\
N205  &$<$ 5.2$\times$10$^{15}$    &1.31$\times$10$^{-2}$       &$>$ 4.4   &3.3         &6.6$\times$10$^{-4}$   &3.9         &0.01\\
      &                            &           &$-$       &            &                       &        &\\
U1503 &1.3$\times$10$^{21}$        &43.6       &2.5       &2.0         &2.2                    &11.4    &0.26\\
      &                            &           &0.4       &            &                       &        &\\
N807  &5.9$\times$10$^{20}$        &43.8       &2.9       &1.5         &2.2                    &8.6     &0.14\\
      &                            &           &0.3       &            &                       &        &\\
N3656 &4.5$\times$10$^{21}$        &101        &2.4       &2.5         &5.1                    &12.5    &0.62\\
      &                            &           &0.1       &            &                       &        &\\
N4476 &$<$ 1.9$\times$10$^{19}$    &4.34       &$>$ 3.4   &$>$ 3.2     &0.22                   &6.8     &0.12\\
      &                            &           &$-$       &            &                       &        &\\
N5666 &2.4$\times$10$^{21}$        &42.5       &2.3       &3.0         &2.1                    &3.7     &0.68\\
      &                            &           &0.1       &            &                       &        &\\
N7468 &8.1$\times$10$^{20}$        &15.9       &2.4       &3.0         &0.80                   &0.76    &0.40\\
      &                            &           &0.1       &            &                       &        &\\
\tableline
\end{tabular}
\begin{tabular}{l}
$^a$T$_D$ is the gas depletion time scale.\\

$^c$Far-infrared spectral index $\alpha$$_{IR}$=$log\left(\frac{S_{60\mu m}}{S_{25\mu m}}\right)/\log\left(\frac{60}{25}\right)$.\\
*All FIR fluxes obtained from (Knapp \etal~1989).\\
\end{tabular}
\end{scriptsize}
\end{table}
\clearpage
\begin{table}
\begin{normalsize}
\caption{\bfseries Molecular Gas Properties}
\begin{tabular}{lccccc}
\\
\tableline\tableline
Galaxy  &M$_{gas}$$^a$         &                        &                       &\\
        &(10$^{8}$ M$_\odot$)  &$\frac{\kappa}{\Omega}$ &M$_{gas}$/M$_{dyn}^a$   &Q\\
\tableline
NGC185........................ &5.6$\times$10$^{-4}$  &$-$  &$-$                &$-$\\
NGC205........................ &2.6$\times$10$^{-3}$  &$-$  &$-$                &$-$\\
UGC1503....................... &25                    &1.4  &0.068$-$0.110      &1.0-1.2\\
NGC807........................ &19                    &1.4  &0.020$-$0.024      &2.5-2.8\\
NGC3656....................... &64                    &2    &0.095$-$0.100      &0.74-0.75\\
NGC4476....................... &1.5                   &2    &0.046$-$0.057      &3.5-4.0\\
NGC5666....................... &7.8                   &2    &0.030$-$0.14       &1.4-3.1\\
NGC7468....................... &0.61                  &$-$  &$\sim$0.004$^b$    &$-$\\
\tableline
\end{tabular}
\begin{tabular}{l}
$^a$ Gas and dynamical masses obtained from Young (2002).  The gas masses\\
include contributions from both hydrogen and helium. The dynamical masses \\
are measured from the gas velocity and radius at the outer edge of the of\\
the CO disk.\\
$^b$In the case of NGC 7468, we have used the radius of the IRAM 30 meter \\
beam (r$\sim$11.5$\arcsec$) since a map is not available.\\
\end{tabular} 
\end{normalsize}
\end{table}
\clearpage
\begin{figure}
\epsscale{0.60}
\plotone{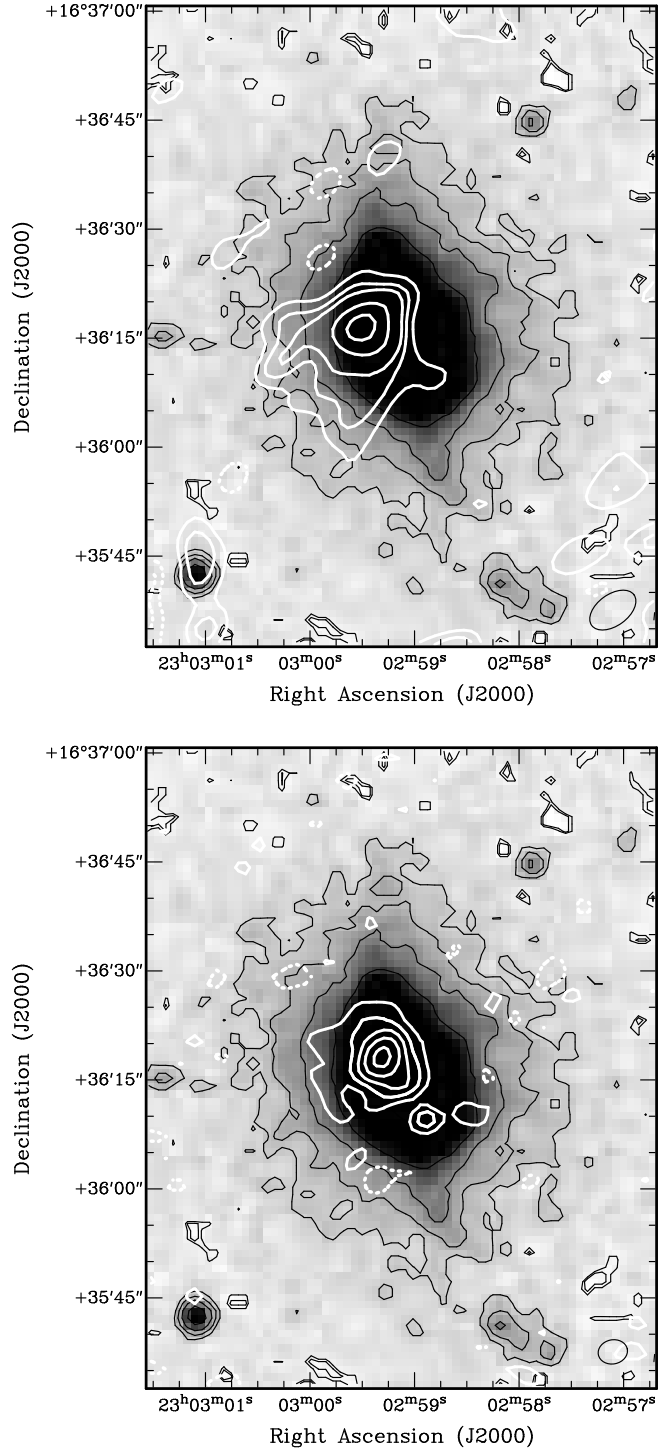}
\caption{\footnotesize NGC 7468. Radio continuum
at 20 cm and 6 cm superposed on an optical image from the 2nd generation digital sky survey.  \textbf{Top}: Heavy (positive) and heavy dashed (negative) white contours show 20 cm radio continuum integrated intensity displayed as multiples of the rms noise (-3, -5, 3, 5, 7, 15, and 20 times 0.1 mJy beam$^{-1}$). Beam size is $7.2'' \times 4.3''$. \textbf{Bottom}: Heavy (positive) and heavy dashed (negative) white contours show 6 cm radio continuum integrated intensity displayed as multiples of the rms noise (-2, -4, 2, 4, 6, 10, and 15 times 0.06 mJy beam$^{-1}$). Beam size is 4.1$''$$\times$3.3$''$. \textbf{Note}: Radio fluxes for this data were published by Cox \& Sparke (2004), but no maps were published.  \label{fig1}}
\end{figure}
\clearpage
\begin{figure}
\epsscale{0.60}
\plotone{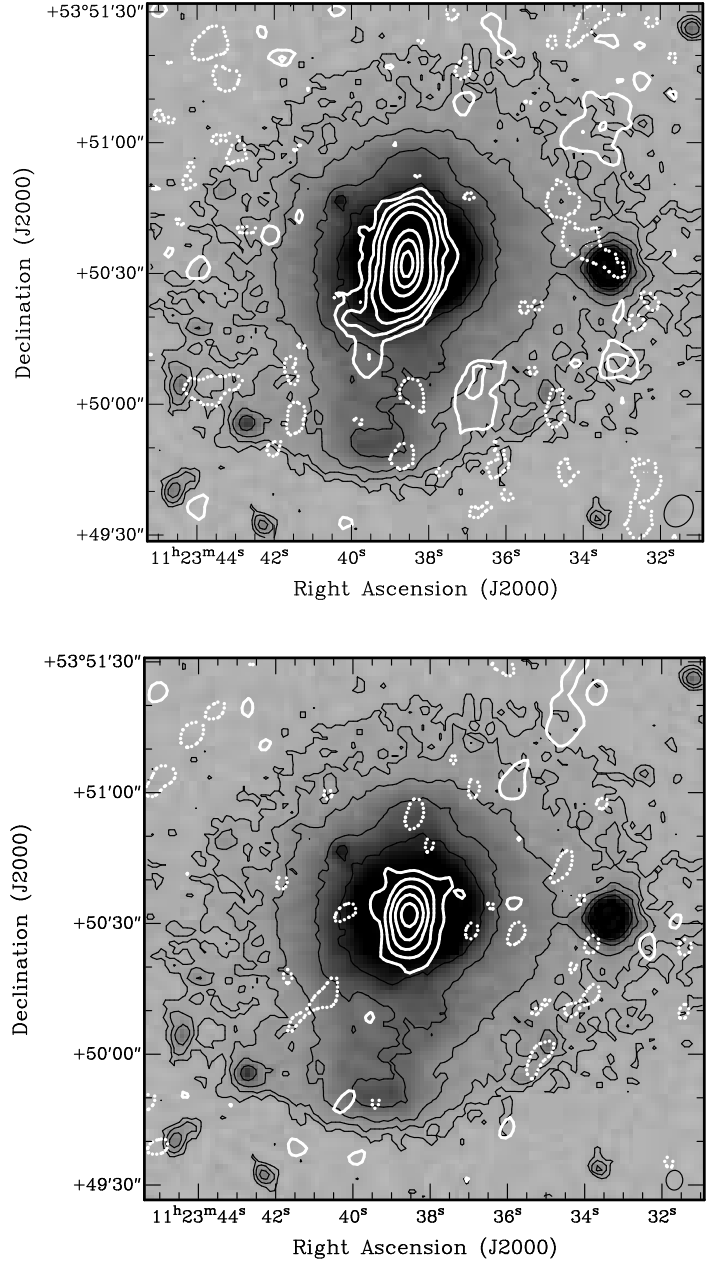}
\caption{\footnotesize NGC 3656.  20 cm radio continuum and 3 mm CO emission superposed on an optical image from the 2nd generation digital sky survey. Figures 2-5 all follow the same pattern with CO on the top and radio continuum on the bottom. \textbf{Top}: CO emission. Heavy (positive) and heavy dashed (negative) white contours show CO integrated intensity displayed in units of -5, -2, 2, 5, 10, 20, 30, 50, 70, and 90 percent of the peak (6.3 Jy beam$^{-1}$ km s$^{-1}$). Beam size is 7.76$''$$\times$6.16$''$. \textbf{Bottom}: Radio continuum.  Heavy (positive) and heavy dashed (negative) white contours show radio continuum integrated intensity displayed as multiples of the rms noise (-3, -10, 3, 10, 30, 80, and 200 times 0.03 mJy beam$^{-1}$). Beam size is 4.5$''$$\times$3.8$''$. Scale in the image is 10$\arcsec=$~2.2 kpc. \label{fig2}}
\end{figure}
\clearpage
\begin{figure}
\epsscale{0.60}
\plotone{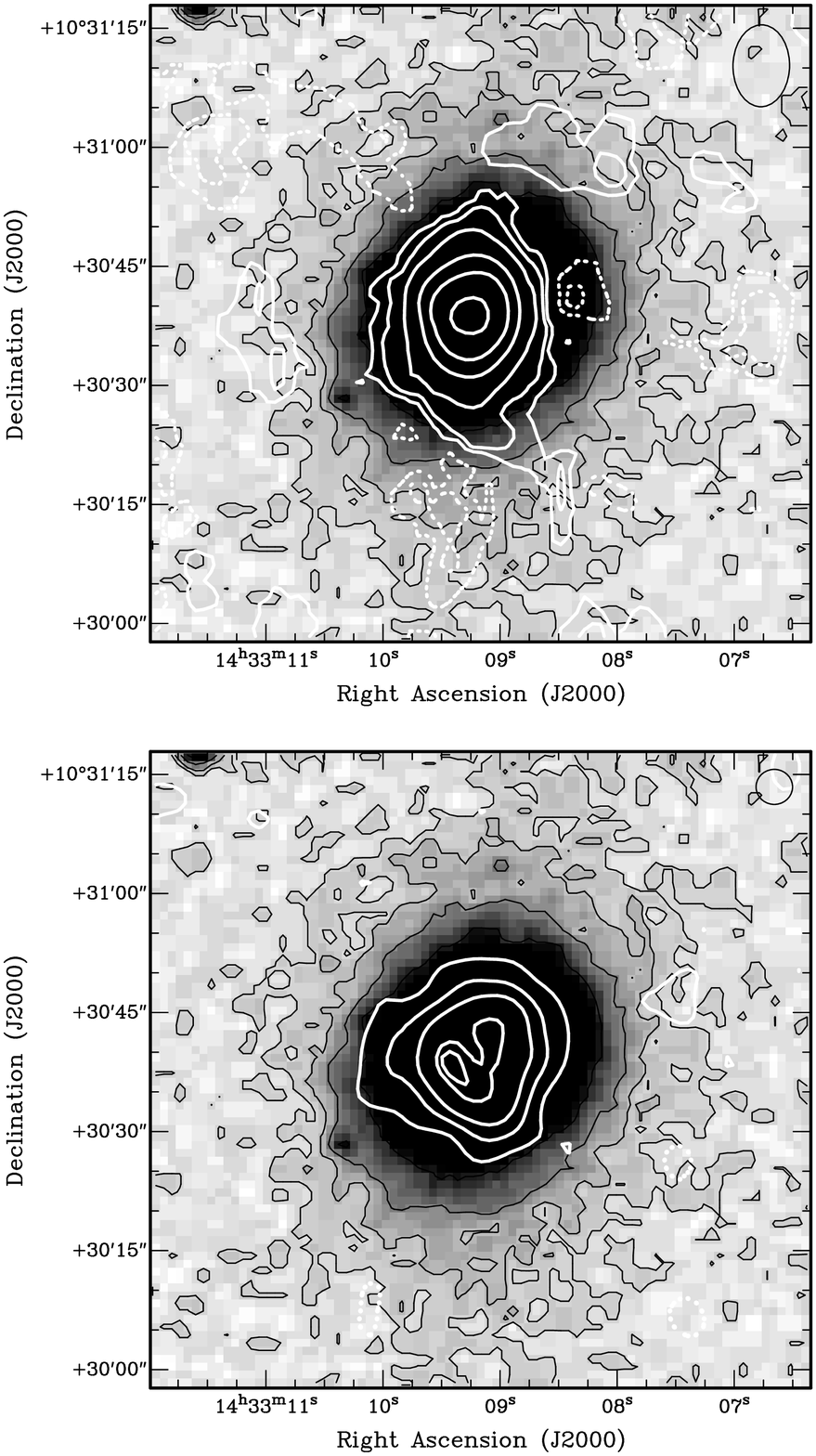}
\caption{\footnotesize NGC 5666.  20 cm radio continuum and 3 mm CO emission superposed on an optical image from the 2nd generation digital sky survey. \textbf{Top}: CO emission.  Heavy (positive) and heavy dashed (negative) white contours show CO integrated intensity displayed in units of -10, -20, 20, 40, 60, 80, and 90 percent of the peak (21.3 Jy beam$^{-1}$ km s$^{-1}$). Beam size is 8.27$''$$\times$6.0$''$.  \textbf{Bottom}: Radio continuum. Heavy (positive) and heavy dashed (negative) white contours show 20 cm radio continuum integrated intensity displayed as multiples of the rms noise (-3, -10, 3, 10, 20, 40, and 45 times 0.04 mJy beam$^{-1}$). Beam size is 4.6$''$$\times$4.4$''$. Scale in the image is 10$\arcsec=$1.7 kpc. \label{fig3}}
\end{figure}
\clearpage
\begin{figure}
\epsscale{0.60}
\plotone{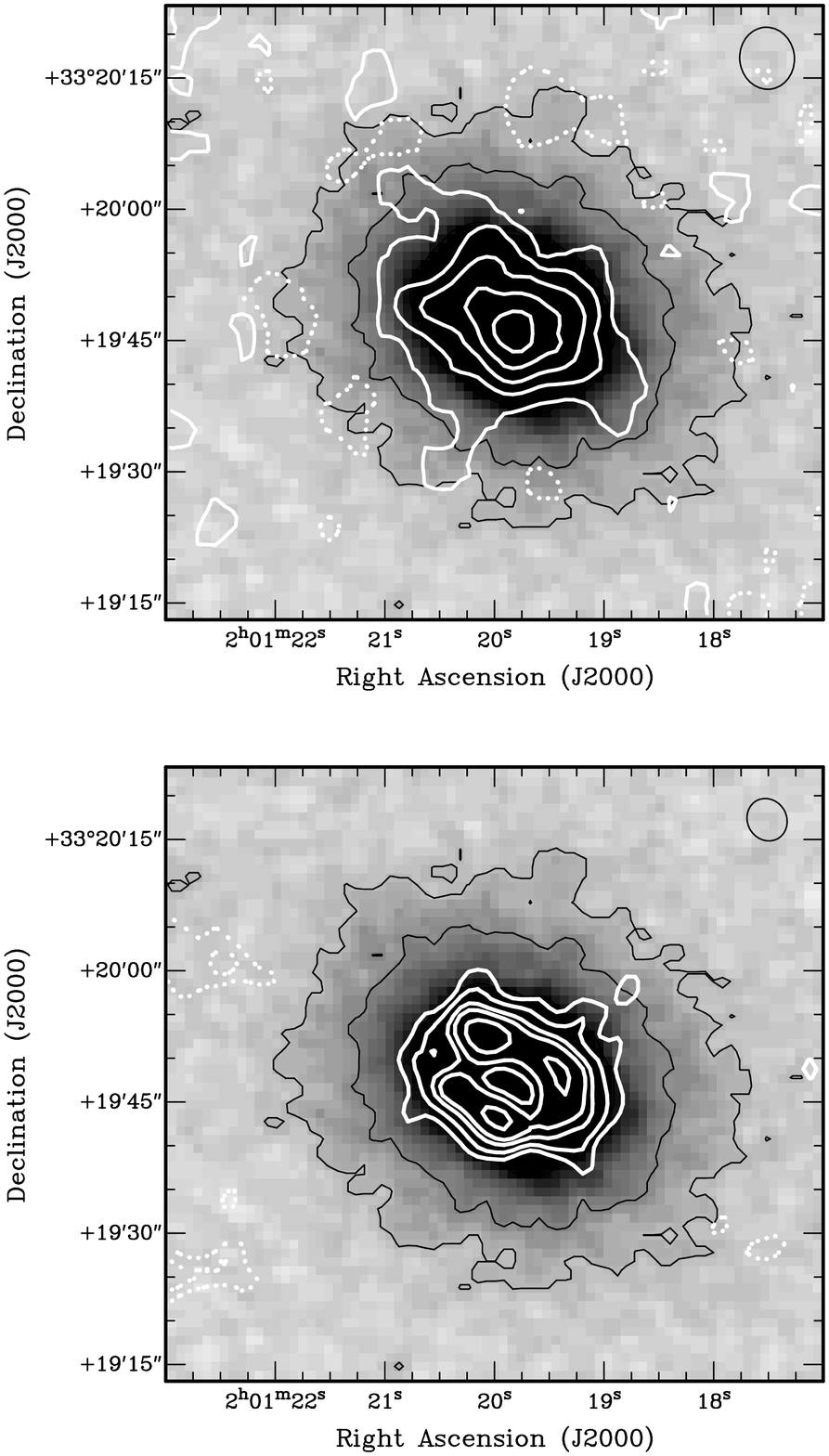}
\caption{\footnotesize UGC 1503.  20 cm radio continuum and 3 mm CO emission superposed on an optical image from the 2nd generation digital sky survey. \textbf{Top}: CO emission.  Heavy (positive) and heavy dashed (negative) white contours show CO integrated intensity displayed in units of -20, -10, 10, 30, 50, 70, and 90 percent of the peak (21.3 Jy beam$^{-1}$ km s$^{-1}$). Beam size is 7.1$''$$\times$6.3$''$.  \textbf{Bottom}: Heavy (positive) and heavy dashed (negative) white contours show radio continuum integrated intensity displayed as multiples of the rms noise (-3, -4, 3, 4, 5, 6 and 8 times 0.03 mJy beam$^{-1}$). Beam size is 5.0$''$$\times$4.5$''$.  Scale in the image is 10$\arcsec=$3.3 kpc. \label{fig4}}
\end{figure}
\clearpage
\begin{figure}
\epsscale{0.60}
\plotone{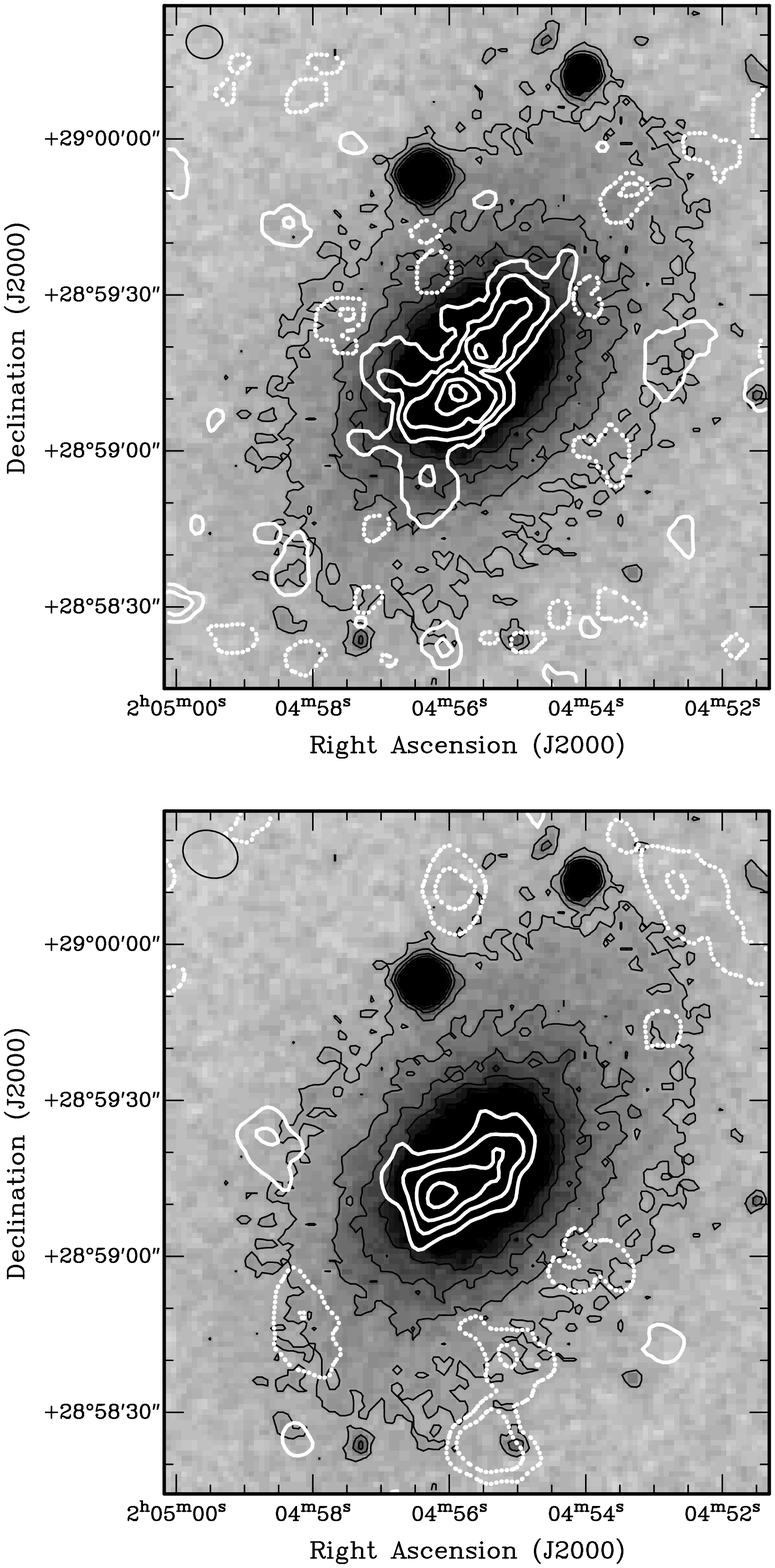}
\caption{\footnotesize NGC 807.  20 cm radio continuum and 3 mm CO emission superposed on an optical image from the 2nd generation digital sky survey. \textbf{Top}: CO emission. Heavy (positive) and heavy dashed (negative) white contours show CO integrated intensity displayed in units of -20, -10, 10, 20, 40, 40, 70, and 90 percent of the peak (3.56 Jy beam$^{-1}$ km s$^{-1}$). Beam size is 7.0$''$$\times$6.3$''$.  \textbf{Bottom}: Heavy (positive) and heavy dashed (negative) white contours show 20 cm radio continuum integrated intensity displayed as multiples of the rms noise (-2, 2, 3, 4, and 5 times 0.05 mJy beam$^{-1}$). Beam size is $10.9'' \times 8.8''$.  Scale in the image is 10$\arcsec=$3.1 kpc. \label{fig5}}
\end{figure}
\clearpage
\begin{figure}
\epsscale{0.60}
\plotone{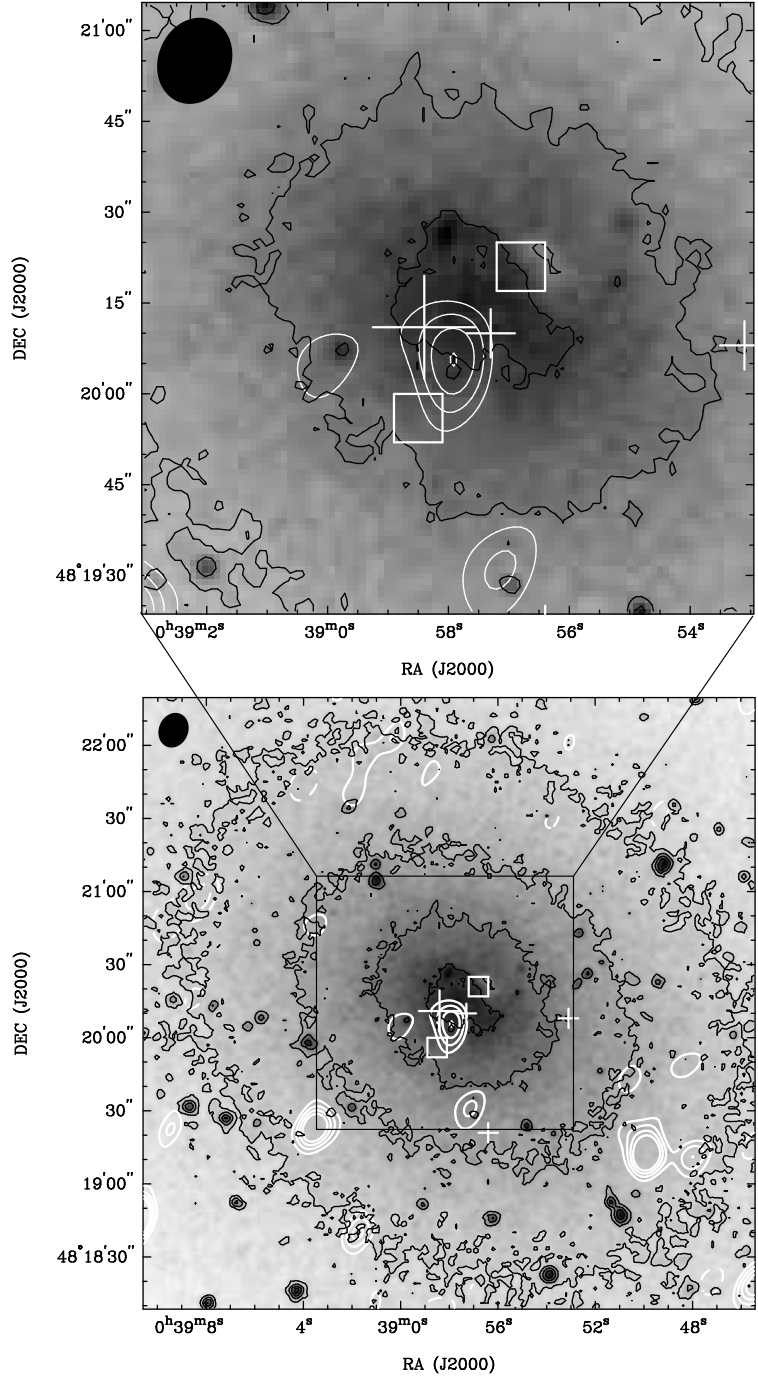}
\caption{\footnotesize NGC 185.  20 cm radio continuum emission at superposed on a 2nd generation digital sky survey red image. White contours show radio continuum integrated intensity displayed as multiples of the rms noise (-4, -2, 2, 3, 4, and 5 times 0.03 mJy beam$^{-1}$). Beam size is $14.6'' \times 12.0''$.  Overlay symbols represent data obtained by Young \& Lo (1997).  The large plus is centered on extended H$_\alpha$ emission, and the diameter of the large plus represents the true diameter of the emission.  The small pluses represent regions where compact H$_\alpha$ emission is present.  The open squares are centered on the positions of CO detections, and these are not to scale.  Scale in the image is 10$\arcmin=$1.8 kpc. \label{fig6}}
\end{figure}
\clearpage
\begin{figure}
\plotone{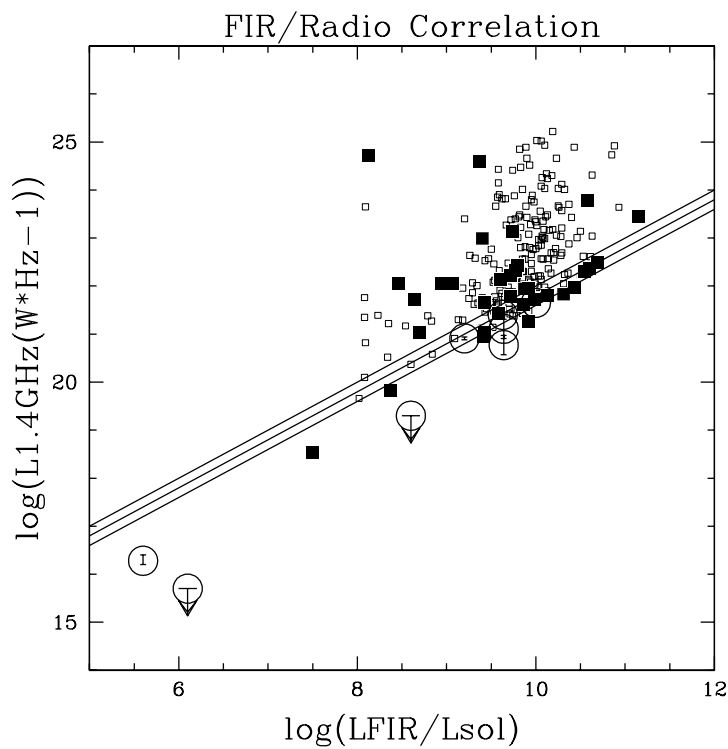}
\caption{\footnotesize Radio/far-IR correlation. The solid line corresponds to a linear relation where $<$q$>$=2.3 and $\sigma_q$=~0.2 for star forming spirals (Condon 1992).  The two black lines represent 1$\sigma$ on either side of the correlation.  The squares represent all early-types detected in radio emission from Condon, Cotton, $\&$ Broderick (2002).  The open squares are FIR upper limits, and the filled squares represent FIR detections.  The black circles correspond to new data presented in this paper. Note that the three galaxies with the smallest amount of radio emission and FIR emission (NGC 185, NGC 205, and NGC 4476) also have the lowest optical luminosities ($\leq~10^9 L_{odot}$) in the sample.  \label{fig7}}
\end{figure}
\end{document}